\documentclass[jcp,aps,showpacs,preprint,byrevtex,amsmath,amssymb,
preprintnumbers,unsortedaddress]{revtex4}

\usepackage{graphicx}

\begin{document}

\preprint{PREPRINT}

\title{Corresponding states law for a generalized Lennard-Jones potential}

\author{P. Orea}
\affiliation{Instituto Mexicano del Petr\'{o}leo, Direcci\'on de 
Investigaci\'on en Transformaci\'on de Hidrocarburos, Eje Central L\'{a}zaro 
C\'{a}rdenas 152, 07730 M\'{e}xico D.F., Mexico.}

\author{A. Romero-Mart\'inez} \affiliation{Instituto Mexicano del Petr\'{o}leo, 
Direcci\'on de Investigaci\'on en Exploraci\'on y Producci\'on,
Eje Central L\'{a}zaro C\'{a}rdenas 152, 07730 M\'{e}xico D.F., Mexico.}

\author{E. Basurto} \affiliation{\'Area de F\'isica de Procesos Irreversibles, 
Divisi\'on de Ciencias B\'asicas e Ingenier\'{i}a, Universidad Aut\'onoma 
Metropolitana-Azcapotzalco, Av. San Pablo 180, 02200 M\'exico 
D.F., Mexico}

\author{C. A. Vargas} \affiliation{\'Area de F\'isica de Procesos 
Irreversibles, Divisi\'on de Ciencias B\'asicas e Ingenier\'{i}a, Universidad 
Aut\'onoma Metropolitana-Azcapotzalco, Av. San Pablo 180, 02200 M\'exico 
D.F., Mexico}

\author{G. Odriozola}
\email{godriozo@azc.uam.mx}
\affiliation{\'Area de F\'isica de Procesos Irreversibles, 
Divisi\'on de Ciencias B\'asicas e Ingenier\'{i}a, Universidad Aut\'onoma 
Metropolitana-Azcapotzalco, Av. San Pablo 180, 02200 M\'exico 
D.F., Mexico}

\date{\today}

\begin{abstract}
It was recently shown that vapor-liquid coexistence densities derived from Mie 
and Yukawa models collapse to define a single master curve when represented 
against the difference between the reduced second virial coefficient at the 
corresponding temperature and that at the critical point. In this work we 
further test this proposal for another generalization of the Lennard-Jones pair 
potential. This is carried out for vapor-liquid coexistence densities, surface 
tension, and vapor pressure, along a temperature window set below the critical 
point. For this purpose we perform molecular dynamics simulations by varying 
the potential softness parameter to produce from very short to intermediate 
attractive ranges. We observed all properties to collapse and yield master 
curves. Moreover, the vapor-liquid curve is found to share the exact shape of 
the Mie and attractive Yukawa. Furthermore, the surface tension and the 
logarithm of the vapor pressure are linear functions of this difference of 
reduced second virial coefficients. 
\end{abstract}

%\pacs{???}

\maketitle

\section{Introduction}

Leaving aside conformal pair potentials (those which are invariant 
by rescaling distance and energy and for which the
corresponding states law is strictly valid~\cite{Pitzer39}), the classic van 
der Waals framework works relatively well for defining master curves for 
thermodynamic properties derived from pair potential shapes of variable 
attractive range~\cite{Okumura00,Dunikov01, 
Reyes08,Galliero09,Blas12,Macdowell12,Chapela13,Lemus13,Singh12} and real 
systems~\cite{Katsonis06,Valadez12,Mulero05,Weiss07,Galliero10,Tang13,Marageh06,
Castellanos04,Ghatee02,Puosi11,Boire13}. This quasiuniversality, 
together with the Lindemann melting criterion~\cite{Ubbelohde}, the 
so-called excess-entropy scaling of liquid's relaxation times and diffusion 
coefficients~\cite{Rosenfeld77}, among other approximate corresponding states 
features~\cite{Young03,Galliero07,Galliero08}, has lead to discover 
that many liquids and solids have an approximate hidden scale invariance, 
implying the existence of isomorph lines in the thermodynamic phase diagram 
along which reduced structure and dynamic properties are invariant to a good 
approximation~\cite{Bacher14,Dyre14}. In other words, the phase diagram becomes 
one-dimensional with regard to several physical properties.    

In line with this scale invariance, quite recently we have observed 
that a slight modification of the extended law of corresponding 
states~\cite{Noro00} is capable of improving the output of the van der Waals 
framework for the Mie and attractive Yukawa expressions 
(which are non-conformal potential functions)~\cite{Orea15}. 
Results have shown that this proposal not only improve the Mie and Yukawa data 
collapse but the obtained master curves were indistinguishable. Hence the 
following questions naturally raise: Is this framework general, leading always 
to data collapses for spherically symmetric potentials? Are the Mie and Yukawa 
the only functional shapes sharing a master curve for the vapor-liquid 
coexistence? Are the vapor-liquid coexistence densities the only properties 
showing this general behavior? This work can be seen as an effort to answer, at 
least partially, these questions. 

The extended law of corresponding states~\cite{Noro00} was derived as an attempt 
to generalize the classic van der Waals principle~\cite{Pitzer39,Guggenheim45} 
to non-conformal potentials, in particular for pair potentials of 
variable attractive range. This, in view that the nature of 
colloidal interactions show such character~\cite{Hunter}. For this purpose, 
Noro and Frenkel suggested including the reduced second virial coefficient, 
$B_2^*$, as an additional independent variable which is clearly linked to the 
attractive range~\cite{Noro00}. This is justified since the value of $B_2^*$ 
at the critical point is frequently close to $-1.5$ for several non-conformal 
potentials~\cite{Vliegenthart00}. Nonetheless, small deviations from this 
particular value have an undesirable impact on the definition of master curves 
for all properties. Hence, we proposed $B_{2s}=B_2^*(T^*)-B_2^*(T^*_c)$ as the 
additional independent variable instead (being $T^*_c$ the dimensionless 
critical temperature)~\cite{Orea15}. 

In this work we are testing the performance of this framework for another 
continuous non-conformal potential shape. This is the so-called Approximate 
Non-Conformal (ANC) potential~\cite{delRio05}, which can be seen as a 
generalization of the Lennard-Jones model or even a 
generalization of a spherically symmetric Kihara~\cite{Kihara54}. The ANC 
expression provides a family of potential functions that 
accurately give the dilute vapor phase properties of several real 
substances~\cite{McLure99,delRio07,delRio13}. The potential 
allows the tuning of its (and attractive range) by 
varying a single parameter, $s$. It has also the advantage of showing an 
analytical second virial coefficient~\cite{JoseA15}. Its spherical symmetry and 
non-conformal character make it appropriate for our purpose. Hence, we perform 
molecular dynamics simulations (using the 
Gromacs package~\cite{gromacs1,gromacs2}) to obtain vapor-liquid coexistence 
densities, surface tension, and vapor pressure for this particular functional 
shape. These data allow us to build master curves for the three properties as 
tentative universal forms. In fact, we corroborate the shape obtained in 
previous work for the vapor-liquid densities~\cite{Orea15}. It is 
also shown that this slight modification of the Noro and Frenkel extended law of 
corresponding states~\cite{Noro00} is capable of producing striking data 
collapses of several properties when varying the potential range, and leads to 
really simple relationships between several properties and $B_{2s}$. Capturing 
a universal behavior (independence of the properties on the details and range 
of the pair-potential function) is important in the soft matter field since 
colloidal interactions strongly depend on the composition of both phases, 
continuous and dispersed~\cite{Hunter}.

The manuscript is structured as follows. After this brief introduction we 
present in section II the model potential and the methods employed for obtaining 
the vapor-liquid coexistence densities, vapor pressure, surface tension, and 
critical properties. Section III shows the raw results and the outcomes from 
both, the classic van der Waals and the extended frameworks. Here we give 
simple expressions for the vapor pressure and the surface tension master 
curves. Finally, section IV presents the more relevant conclusions. 

\section{Model and method}

\begin{figure}
\resizebox{0.55\textwidth}{!}{\includegraphics{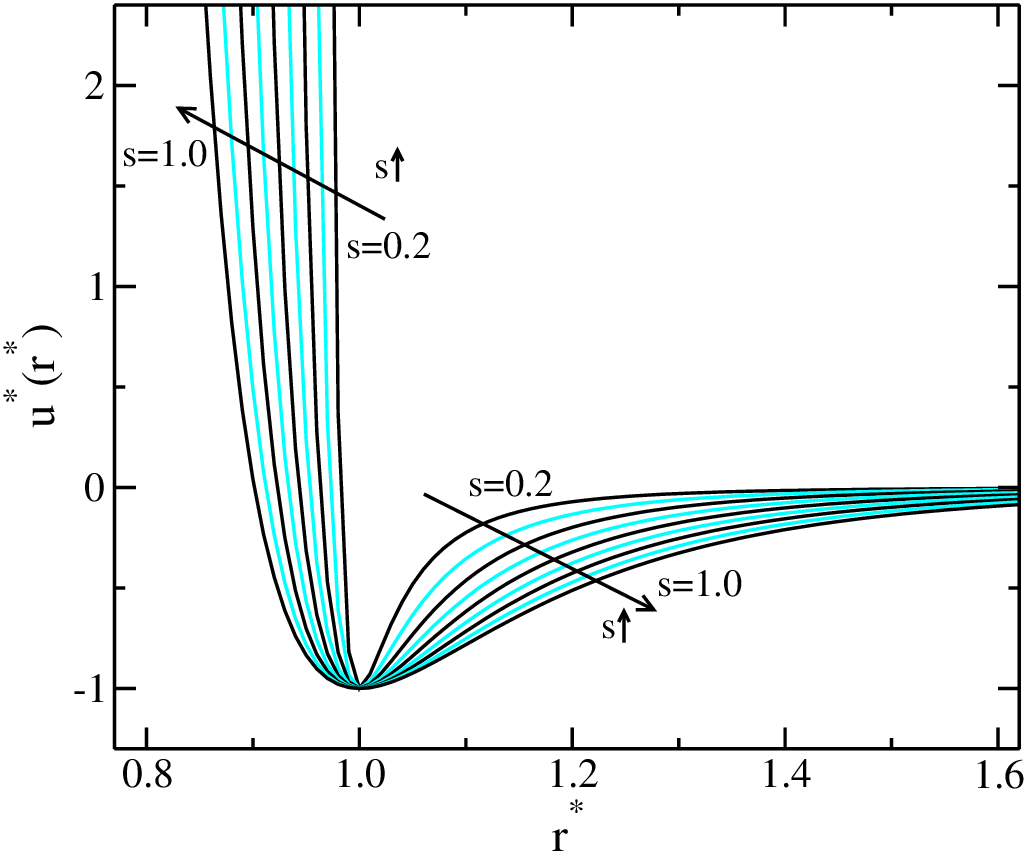}}
\caption{\label{pot} Pair potential $u^*(r^*)$ as a function of the 
softness parameter, $s$. Parameter $s$ increases from 0.2 to 1.0 
in steps of 0.1. Arrows point to the direction of increasing $s$. A larger $s$ 
value leads to a softer curve and a longer interaction range. }
\end{figure}

In reduced units ($u^*=u/\varepsilon$, being $\varepsilon$ the potential well 
depth) the ANC pair potential is given by~\cite{delRio98}
\begin{equation}\label{ANC}
u^*(r^*) = \left[ \frac{1-a}{\xi(r^*) - a} \right]^{12} - 
2\left[ \frac{1-a}{\xi(r^*) - a}\right]^{6} {,}
\end{equation}
where $a=0.09574$ is a constant, $r^*$ is the dimensionless distance $r^* 
\equiv r/r_m$,  $r_m$ is the distance at which the potential reaches its 
minimum,  
\begin{equation}
\xi(r^*) = \left ( \frac{r^{*3}-1}{s}+1\right ) ^{1/3},
\nonumber
\end{equation}
and $s$ is the so called softness parameter. For $s=1$, 
equation~\ref{ANC} produces a spherically symmetric Kihara with hard-core 
diameter $a$. On the other hand, for the fixed value of $a$, 
$s=1.13$ approaches the Lennard-Jones interaction
~\cite{delRio05,delRio07} (the ANC leads to the exact Lennard-Jones 
interaction for $a=0$ and $s=1$). The effect of $s$ on the shape 
of the potential is shown in figure~\ref{pot}. Note that expression~\ref{ANC} 
works properly for inter-particle distances above $r^*_s=(s(a^3-1)+1)^{1/3}$. 
For $r^* \le r^*_s$ we are setting $u^*(r^*)=u^*(r^*_s)$, being for all cases 
$u^*(r^*_s)$ a very large positive value (the value of $u^*(r^*)$ for 
$r^*<r^*_s$ is indeed irrelevant). Along the manuscript we are using 
$\rho^*=\rho r_m^3$ as unit of density (being $\rho$ the number density), 
$T^*=k_BT/\varepsilon$ (being $k_B$ the Boltzmann constant), $\gamma^*=\gamma 
r_m^2/\varepsilon$, $P^*=P r_m^3/\varepsilon$, 
and $t^*=t\sqrt{\varepsilon/mr_m^2}$ ($m$ is the particle 
mass).

This potential is tabulated to be used as input for the Gromacs 
molecular dynamics package~\cite{gromacs1,gromacs2}. We are performing $NVT$ 
replica exchange simulations expanding the ensemble 
in temperature~\cite{Lyubartsev92,Marinari92}. The velocity rescale algorithm 
is employed as thermostat~\cite{Bussi07}. The step time is set to $dt^*= 0.001$ 
 (except for the case with $s=0.2$ which is $dt^*= 0.0005$). The 
cutoff distance, $r_c$, for tables and neighbor list searching 
is such that $|u^*(r_c)|<10^{-4}$ for all cases. We are considering a simulation 
cell having a rectangular parallelepiped shape with $L_x=L_y=9 r_m$ and $L_z=30 
r_m$ for $s<0.7$ and with $L_x=L_y=10 r_m$ for the cases having $s \ge 0.7$, 
such that a liquid slab is kept at the parallelepiped center surrounded by a 
vapor phase~\cite{Chapela77}. We employed these two box sizes due to a twofold 
purpose, to safely increase the cutoff as demanded by the above given 
conditions, and to decrease the overall box density (the critical density 
diminishes with $s$). Initially, we randomly place $N=1200$ particles inside the 
central slab and let the system relax. Periodic boundary conditions are set for 
the three orthogonal directions. The trajectories expand a total time of 
$t^*=2\times10^4$. Eight replicas are considered for each value of $s$ and 
different temperatures. Temperatures are fixed following a geometrical 
decreasing trend, such that the highest temperature is close to (but below) the 
critical point. The geometrical factor is chosen to obtain swap acceptance rates 
above $0.1$. 

\begin{figure}
\resizebox{0.55\textwidth}{!}{\includegraphics{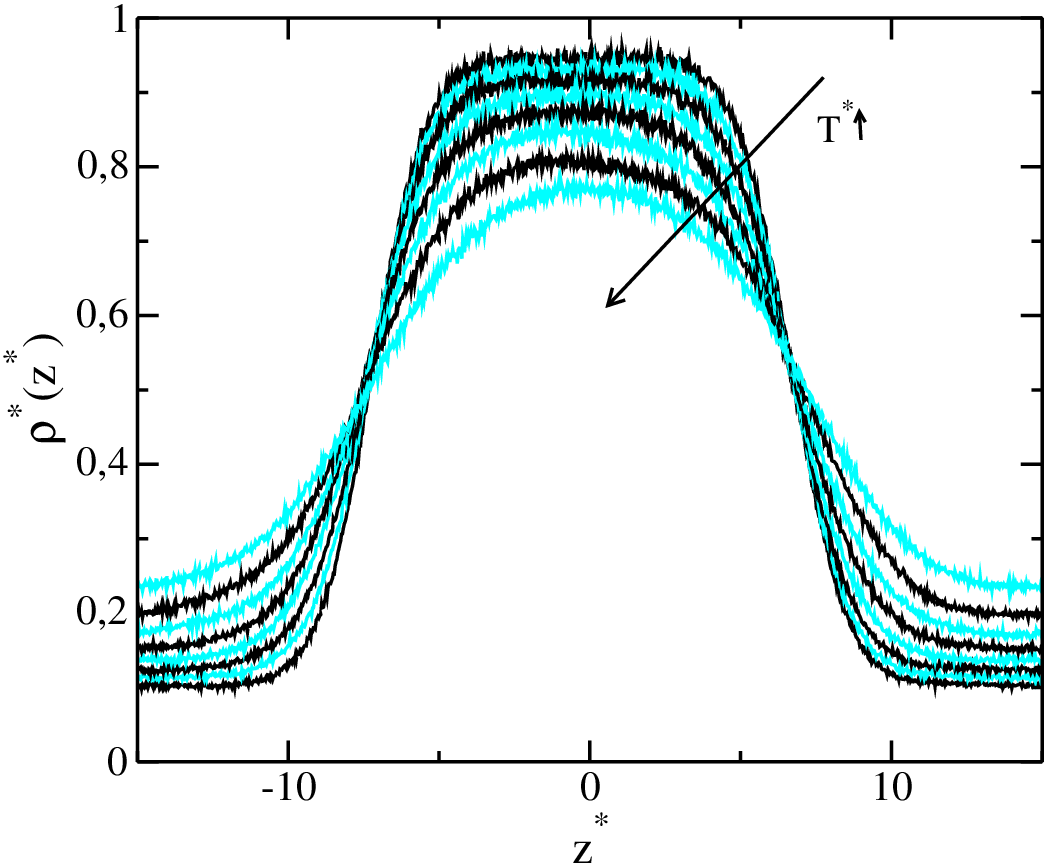}}
\caption{\label{perfiles} Density profiles along the largest cell side (normal 
to the interfaces) for the shortest range case, $s=0.2$. Different curves 
correspond to different temperatures. The arrow points to the direction of 
increasing temperature. }
\end{figure}

The trajectories are then analyzed by means of a simple home made program code 
which produces the density profiles (fixing the system center of mass at the 
parallelepiped geometrical center) and discards the first steps where energy has 
not reached a clear plateau. The output is shown in figure~\ref{perfiles} for 
$s=0.2$ and for all different temperatures. It is worth mentioning that the 
difficulty for getting acceptable profiles increases with decreasing the 
potential range. As can be seen in figure~\ref{pot} the potential shape is quite 
sharp for $s=0.2$. Such a short range implies low coexistence temperatures, long 
living bonds, a slow dynamics, and a more likely crystallization. Hence, we are 
showing the most difficult case. The profiles are then employed to obtain the 
vapor and liquid densities from the fully developed bulk regions. Note that 
there is not a clear liquid bulk region for the largest temperature shown in 
figure~\ref{perfiles}. In such a case the point is discarded for determining the 
critical properties. The critical density $\rho^*_c$ and temperature $T^*_c$ are 
obtained by considering the effective critical exponent $\beta_e=0.325$ and from 
the law of rectilinear diameters. The critical pressure is taken from a linear 
extrapolation of the logarithm of the vapor pressure against the reciprocal 
temperature towards $1/T^*_c$.

The pressure tensor is obtained from the virial expression~\cite{Frenkel}. In 
turn, the surface tension can be found by means of
\begin{equation}\label{st}
\gamma=\frac{L_z}{2}  \left \{ \langle P_{zz} \rangle -1/2 (\langle P_{xx} \rangle+\langle P_{yy} \rangle)   \right \}
\end{equation}
where $P_{ii}$ ($i=$ $x$, $y$, $z$) are the diagonal components. The factor 2 
is due to the existence of two interfaces. Our output is identical to the one 
obtained from the Gromacs tools (g\_energy) once the factor of 2 is accounted 
for. The vapor pressure, $P$, is taken as the normal to the interfaces, 
$P_{zz}$. 

\begin{figure}
\resizebox{0.65\textwidth}{!}{\includegraphics{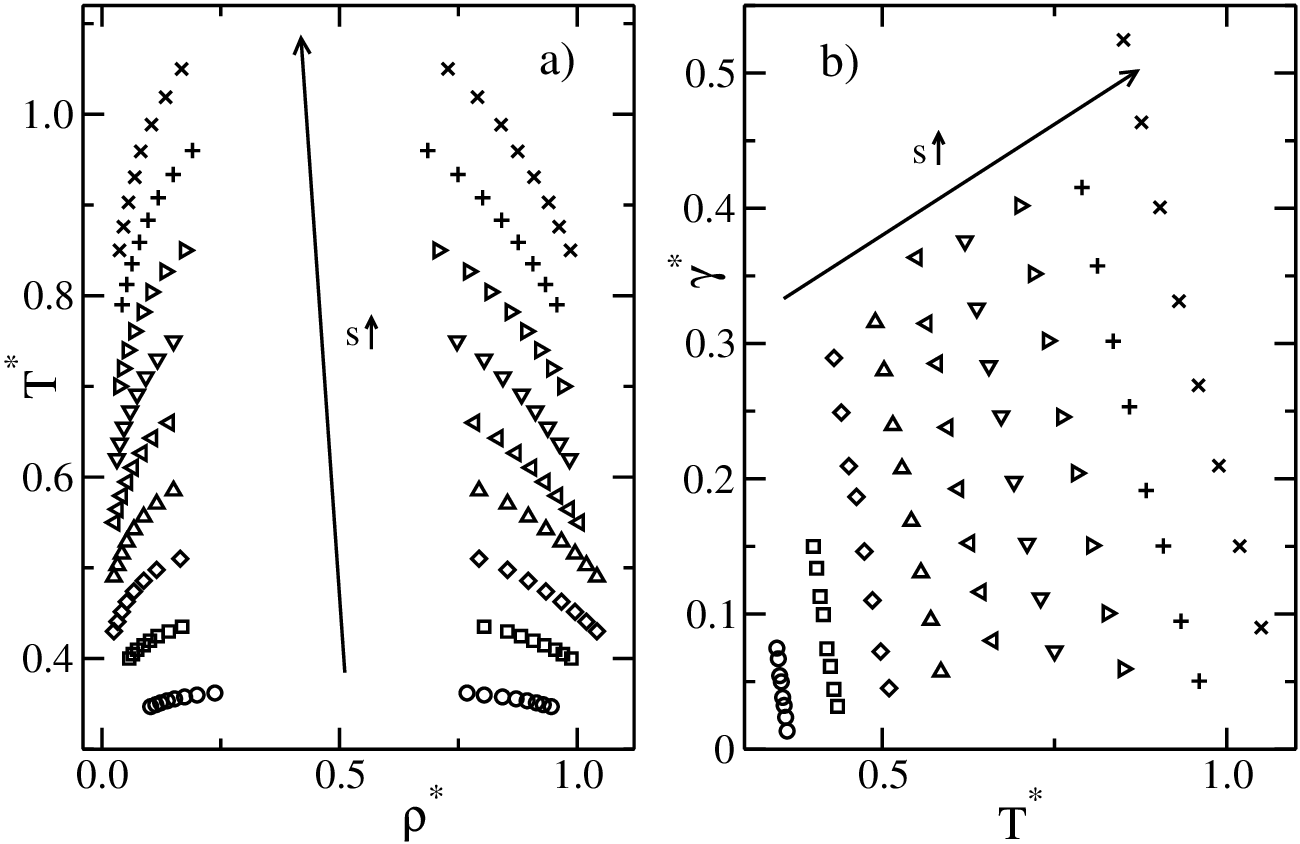}}
\caption{\label{coexist-tens} a) Vapor-liquid coexistence density curves for 
the ANC potential with varying $s$. Circles, squares, diamonds, triangles-up, 
triangles-left, triangles-down, triangles-right, plus symbols, and crosses 
correspond to $s=0.2, 0.3, 0.4, 0.5, 0.6, 0.7, 0.8, 0.9,$ and $1.0$, 
respectively. b) The corresponding dimensionless surface tension (the estimated 
error is always below $5\%$). Arrows indicate the direction of increasing $s$.  
}
\end{figure}

In previous work we have defined $B_{2s}^*(T^*)=B_2^*(T^*)-B_2^*(T^*_c)+cst.$ 
where $B_2^*(T^*)=3B_2(T^*)/(2\pi \sigma^{*3}_{eff}(T^*))$,
\begin{equation}
B_2(T^*)=2 \pi \int_0^\infty  r^{*2} [1- e^{-u^*(r^*)/T^*}] dr^*,
\end{equation}
\begin{equation}
\sigma^*_{eff}(T^*)= \int_0^\infty  [1- e^{-u'^*(r^*)/T^*}] dr^*,
\end{equation}
$u'^*(r^*)=u^*(r^*)+1$ for $r^*<1$ and  $u'(r^*)=0$ 
otherwise~\cite{Andersen71,Barker76}. Note that $2\pi \sigma^{*3}_{eff}(T^*)/3$ 
is the second virial coefficient of hard spheres with a hard core diameter of 
$\sigma^*_{eff}$. There, we have employed $cst.=-1.5$ to gain consistency with 
previous works. Now we are setting $cst.=0$ to gain simplicity of the 
expressions. Furthermore, the difference between the reduced second virial 
coefficient at $T^*$ and $T^*_c$ would be our measure of the attractive range. 
This way, the fitted expressions for the liquid and vapor branches of the 
coexistence are given by
\begin{equation} \label{liquid}
B^*_{2s}(\rho^*)=-0.475(\rho^*-1)^{3.1}
\end{equation}
and
\begin{equation}\label{vapor}
 B^*_{2s}(\rho^*)=-(b\rho^{*3}+c\rho^{*1/2})^{-1}+(b+c)^{-1} 
\end{equation}
with $b=75.1$ and $c=3.71$, respectively. These 
expressions were obtained by means of a trial and error procedure 
and correspond to both, the Mie and Yukawa potentials. 

\section{Results}

As explained, from the fully developed liquid and vapor bulk phases we obtain 
coexistence densities. These are presented in figure~\ref{coexist-tens} a). 
There are eight points per curve, each one corresponding to a different value of 
the softness parameter, $s$. Each of the eight points corresponding, in turn, to 
different temperatures (we are setting eight replicas). The embedded arrow in 
the left panel shows the increasing $s$ direction. Curves shift to larger 
temperatures when increasing the potential range, as usual (more kinetic energy 
is required to produce the vapor phase). Note also that the temperature window 
is strongly reduced for the strong short range cases. This is done on purpose to 
avoid the formation of a crystal phase (a short range potential combined with a 
monodisperse central-core enhance crystallization). Also, the small 
temperatures lead to long lived bonds which yield a slow system dynamics. Both 
issues make it difficult to obtain coexistence vapor-liquid curves for strong 
short range interactions. The data are in good agreement with the Gibbs ensemble 
Monte Carlo simulations reported in reference~\cite{delRio07}. The $s$ range in 
their study is $0.6 \le s \le 1.2$. For $s=0.5$ our data are in good agreement 
with those reported in reference~\cite{delRio05}. To the best of our knowledge, 
there are no vapor-liquid coexistence data reported in the literature for 
$s<0.5$.   

\begin{figure}
\resizebox{0.65\textwidth}{!}{\includegraphics{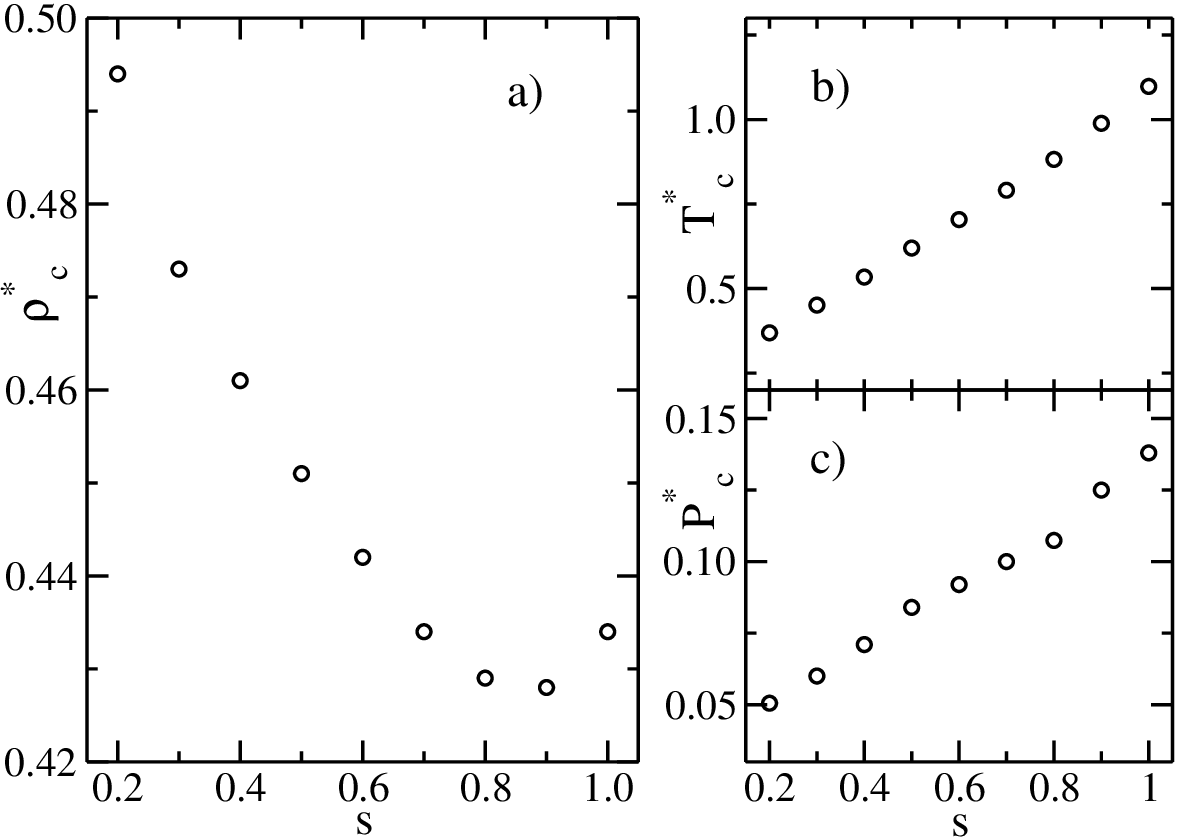}}
\caption{\label{criticos} Critical properties as a function of $s$. a) 
Dimensionless critical density, b) temperature, and c) vapor pressure. Estimated 
errors for density, temperature, and vapor pressure are always below $4\%$, 
$2\%$, and $4\%$, respectively. Numerical values 
are given in table \ref{table1}. }
\end{figure}

\begingroup
\squeezetable

\begin{table}
\caption{Dimensionless critical properties of the ANC potential as a function 
of its softness, $s$. Estimated errors are always lower than 2$\%$, 4$\%$, 
and 4$\%$, for $T^*$, $\rho_c^*$, and $P_c^*$, respectively. 
$Z_c$ is obtained from $P_c^*/(T_c^*\rho_c^*)$ and thus its error 
is lower than $10\%$. }

\label{table1}

\begin{tabular}{ccccccc}
\hline\hline
\hspace{1.0cm}$s$ \hspace{0.5cm} &\hspace{0.5cm} $T_c^*$\hspace{0.5cm} & 
$\rho_c^*$\hspace{0.5cm}
& \hspace{0.5cm} $P_c^*$\hspace{0.5cm} & \hspace{0.5cm} $Z_c$\hspace{0.5cm} & 
\hspace{0.5cm} $B^*_{2}(T_c^*)$ \hspace{1.0cm}  \\

\hline  
  0.20  & 0.369  & 0.494  & 0.051  & 0.277  & -1.309  \\
  0.30  & 0.451  & 0.473  & 0.060  & 0.281  & -1.331  \\
  0.40  & 0.534  & 0.461  & 0.071  & 0.288  & -1.362  \\
  0.50  & 0.620  & 0.451  & 0.084  & 0.300  & -1.387  \\
  0.60  & 0.704  & 0.442  & 0.092  & 0.296  & -1.438  \\
  0.70  & 0.791  & 0.434  & 0.100  & 0.291  & -1.490  \\
  0.80  & 0.882  & 0.429  & 0.107  & 0.284  & -1.541  \\
  0.90  & 0.989  & 0.428  & 0.125  & 0.295  & -1.559  \\
  1.00  & 1.098  & 0.434  & 0.138  & 0.290  & -1.597  \\

\hline

\end{tabular}
\end{table}

\endgroup
                                             
Dimensionless surface tension data are shown in figure~\ref{coexist-tens} b). 
Different symbols are used for different values of $s$, in correspondence with 
panel a). The arrow points along the increasing $s$ direction. As expected, the 
values increase with decreasing temperature and tend to zero as approaching the 
critical temperature. Also, curves shift to the right with increasing $s$, in 
correspondence to the vapor-liquid coexistence densities. Data for $s=0.5$ 
agree well with those given in reference~\cite{delRio05}. For $s>0.5$ our data 
are well above theirs, most probably due to the use of an insufficiently large 
cutoff, as the authors explain in reference~\cite{delRio07}. Indeed, the 
differences between our data and those reported in reference~\cite{delRio05} 
increase with increasing $s$ confirming this claim. Hence, our surface tension 
data for $s>0.5$ can be considered as the first clean ones given in the 
literature. In addition, the surface tension for the region $0.2 \le s \le 0.4$ 
is technically difficult to access. Hence, and as for the coexistence, there are 
no previously reported surface tension data for this region. 

The monotonic and practically linear trend of the critical temperature with $s$ 
is shown in figure~\ref{criticos} b). Values agree with previously reported 
data~\cite{delRio07} in the interval $0.6 \le s \le 1.0$. Also, the critical 
vapor pressure shows a linear behavior with $s$, figure~\ref{criticos} c). Here 
the general trend agrees with the trend observed by del R\'io et. 
al~\cite{delRio07}. Our values are, however, somewhat smaller than theirs. As 
expected, both properties increase with the potential range. Conversely, the 
critical density shows a more complex behavior, as it is shown in the main 
panel of figure~\ref{criticos}. It probably shows a minimum close to $s=0.9$, 
or simply shift from a clear decay for $s\le0.6$ to a plateau at the interval 
$0.8\le s \le 1.0$. The uncertainty of the data does not allow us to discern one 
scenario from the other. Nonetheless, our data show a better defined trend than 
those reported elsewhere~\cite{delRio05,delRio07}. We get a plateau for the 
critical compressibility, $Z^*_c=P^*_c/(T^*_c \rho_c)$, for $s>0.4$ and a 
probable slight decrease for $s<0.4$ (see table~\ref{table1}). 
Again, the uncertainty of the data hinders the trend which seems to be in line 
with the one reported in ref.~\cite{delRio07}. Finally, the reduced second 
virial coefficient at the critical temperature $B_2^*(T^*_c)$ is an ever 
decreasing function of $s$ in the studied interval. It decays from $-1.309$ to 
$-1.597$ (see table~\ref{table1}).  

\begin{figure}
\resizebox{0.65\textwidth}{!}{\includegraphics{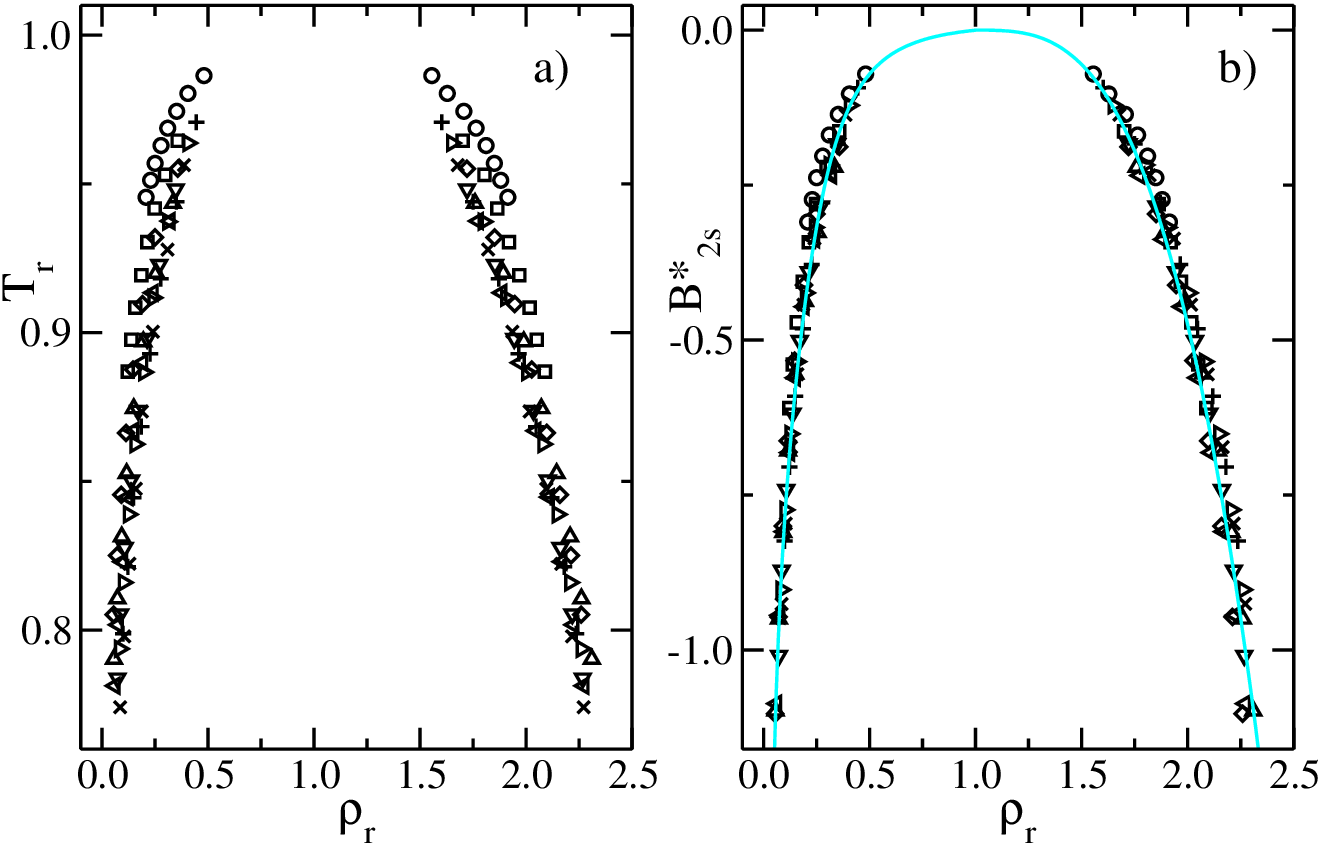}}
\caption{\label{coexist-extlaw} Vapor-liquid coexistence for the ANC potential 
with varying $s$. Different symbols are employed for different $s$ values, in 
correspondence with figure~\ref{coexist-tens}. a) $T_r-\rho_r$ chart as 
following the van der Waals principle. b) $B_{2s}^*-\rho_r$ chart. This last 
chart includes the master curve fit to the Mie and Yukawa potential as a light 
(cyan) line (equations~\ref{liquid} and~\ref{vapor}). }
\end{figure}

Up to this point, we have shown raw data obtained for some 
vapor-liquid coexistence and surface properties for the ANC potential by varying 
$s$ (see table~\ref{table-sup}). From here on, we are 
comparing the outputs from the van der Waals principle to those of the extended 
framework with $B^*_{2s}$. We are omitting plots against $B^*_{2}$ due to the 
$B^*_{2}(T^*_c)$ dependence on $s$. Figure~\ref{coexist-extlaw} shows the 
results of applying the van der Waals 
(left panel) and extended (right panel) frameworks to the vapor-liquid 
coexistence. Reduced properties are given by $T_r=T^*/T^*_c$ and 
$\rho_r=\rho^*/\rho^*_c$. It is observed how the relatively good data collapse 
on the left panel is improved on the right one. This was already pointed out for 
the Mie and Yukawa potentials~\cite{Orea15}. Moreover, we are including the 
master curve obtained from these potentials as a light (cyan) line which, as can 
be seen, can be considered a fit to these new data. So, Mie, Yukawa, and ANC 
lead to the same coexistence density master curve, which imply a correspondence 
between their attractive ranges irrespective of their different shape. This 
result is encouraging since, letting aside the square-well potential 
case~\cite{Orea15}, expressions~\ref{liquid} and~\ref{vapor} seem general. 
Nonetheless, further testing is needed. 

\begin{figure}
\resizebox{0.65\textwidth}{!}{\includegraphics{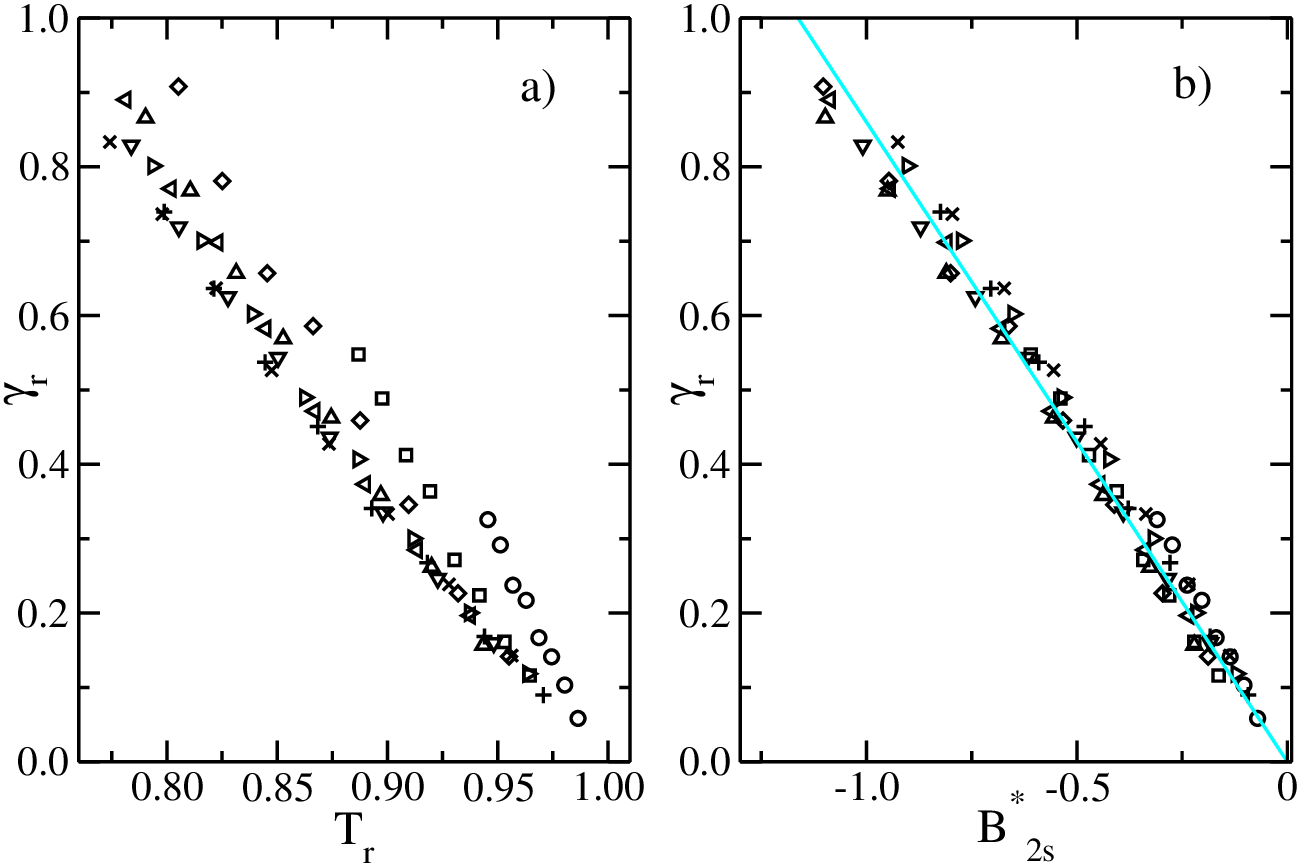}}
\caption{\label{tension-extlaw} Surface tension for the ANC potential with 
varying $s$. Different symbols are employed for different $s$ values, in 
correspondence with figure \ref{coexist-tens}. a) $\gamma_r-T_r$ chart as 
following the van der Waals principle. b) $\gamma_r-B_{2s}^*$ chart. The linear 
fit $\gamma_r=-0.86 B^*_{2s}$ is shown as a light (cyan) line.  }
\end{figure}

We now focus our attention on how the corresponding states frameworks behave 
when dealing with properties such as the surface tension and vapor pressure. The 
left panel of figure~\ref{tension-extlaw} shows a $\gamma_r-T_r$ chart with 
$\gamma_r=\gamma^*/(\rho^*_c T_c^{*2/3})$. Likewise, figure~\ref{tension-extlaw} 
b) shows $\gamma_r$ as a function of $B^*_{2s}$. Again, there is a gain of the 
data collapse when employing $B^*_{2s}$ instead of $T_r$ as the independent 
variable. In the left panel, circles, squares, and probably diamonds, are far 
from where the other points concentrate. This series corresponds 
to $s=0.2$, 0.3, and 0.4, respectively. So, the scattering of the 
curves increases with decreasing $s$. We may say that the collapse appears only 
for $s>0.4$ under the van der Waals framework. This situation is 
not observed in the right-side panel, where all curves seem to obey the 
corresponding states principle. In addition, in base of the estimated errors, we 
cannot discard a linear behavior for the general shape of the obtained master 
curve. This line has a slope of $-0.86$ and zero $y$-intercept. 

\begin{figure}
\resizebox{0.65\textwidth}{!}{\includegraphics{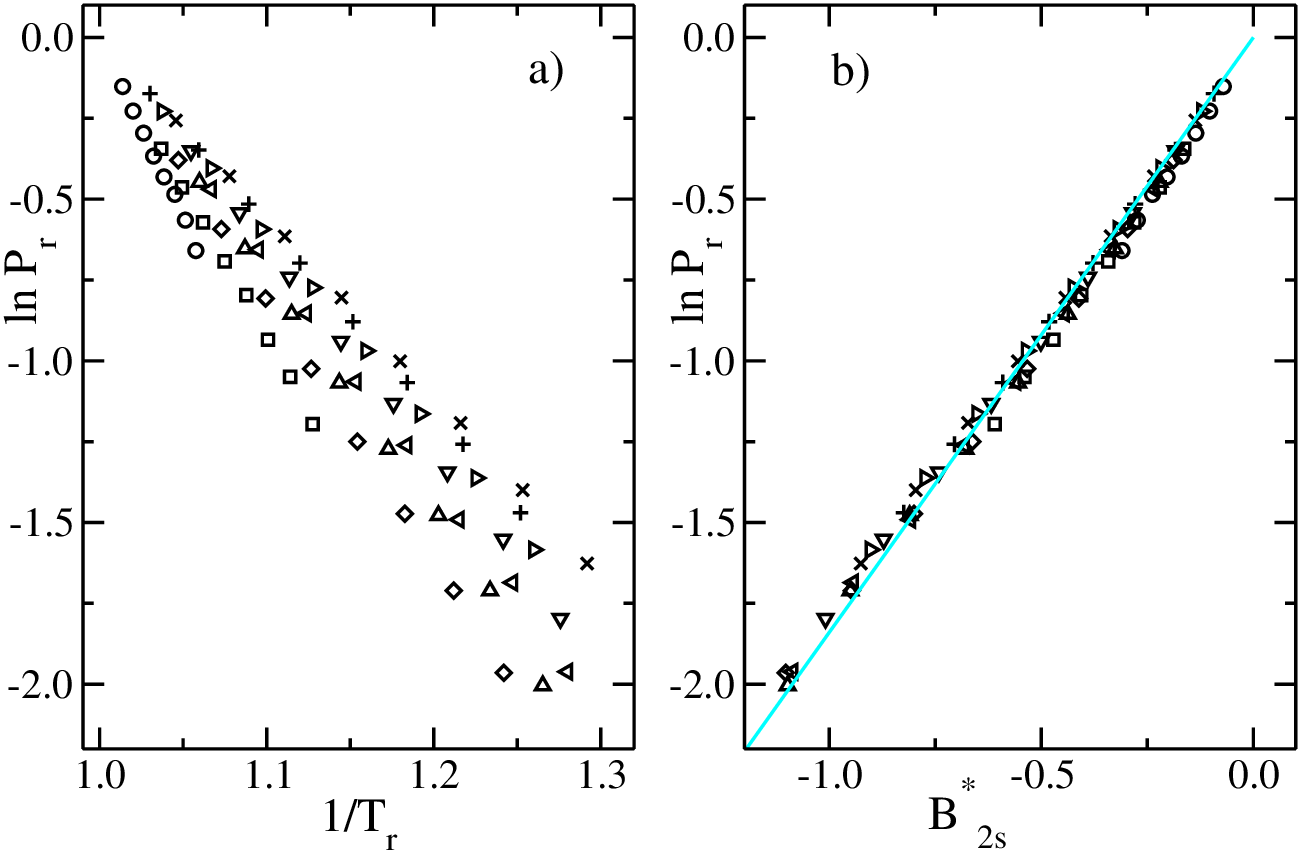}}
\caption{\label{presion-extlaw} Logarithm of the vapor pressure for the ANC 
potential with varying $s$. Different symbols are employed for different $s$ 
values, in correspondence with figure~\ref{coexist-tens}. a) As a function of 
$1/T_r$ as following the van der Waals principle. b) As a function of 
$B_{2s}^*$. The linear fit $\ln P_r=1.84 B^*_{2s}$ is shown as a light (cyan) 
line. }
\end{figure}

Finally, we show the logarithm of the reduced vapor pressure, ln 
$P_r=$ ln $(P^*/P^*_c)$, as a function of $1/T_r$ in 
figure~\ref{presion-extlaw} a) and as a function of $B^*_{2s}$ in 
figure~\ref{presion-extlaw} b). Well defined straight lines are defined in the 
left panel. Their slopes (absolute value), however, 
decrease monotonically with increasing $s$ for all the studied interval. So, a 
master curve cannot be defined. This picture changes when considering the 
extended framework where a single curve appears. Furthermore, the curve turns 
linear when directly plotted against $B^*_{2s}$. The fitted expression, shown as 
a light line, reads $\ln P_r=1.84 B^*_{2s}$. All data lie on this curve when 
taking into account their corresponding error bars (not shown to gain 
clarity).

\section{Conclusions}

We have reported the vapor-liquid coexistence density, vapor pressure, and 
surface tension for the ANC potential (expression~\ref{ANC}) of variable range. 
Its softness parameter, $s$, has been set in the range $0.2 \le 
s \le 1.0$ at intervals of $0.1$. This was done by means of molecular dynamics 
simulations (Gromacs package) in the $NVT$ ensemble. For this purpose, a 
rectangular parallelepiped cell was employed having a liquid slab and two 
vapor-liquid interfaces. Our data agree well with previously reported data for 
the coexisting vapor-liquid densities in the interval $0.5 \le s \le 1.0$ and 
for the surface tension with $s=0.5$. All properties in the interval $0.2 \le s 
\le 0.4$ are reported for the first time. These short range cases are 
technically difficult to access with simulations. We have found a 
linear relationship between $T_c$, $P_c$ and $s$. Conversely, $\rho_c$ is not a 
linear function of $s$. These trends were not clear from previous 
results.  

The van der Waals and extended law frameworks were applied to the obtained 
data. This was carried out for vapor-liquid coexistence densities, surface 
tension, and vapor pressure. For all cases, a much better data collapse is 
observed when using $B^*_{2s}$ as an independent variable instead of the reduced 
temperature. Furthermore, the obtained coexistence density master curve is 
practically the same we have found for the Mie and the Yukawa potentials. 
Finally, the master curves found for surface tension and vapor pressure are 
strikingly simple and tentatively universal. In view of the presented results, 
we expect the master curves for these properties to hold for the Mie and Yukawa 
potential. We also expect other spherically symmetric pair potentials to behave 
in line with the ANC, Mie, and Yukawa fluids. This would imply a correspondence 
between their attractive ranges irrespective of their different shape.

\section{Acknowledgments}

PO and AR thank the Instituto Mexicano del Petr\'{o}leo for financial support 
(Project No D.61017). PO also thanks the IMP Project No Y.60013. GO thanks 
CONACyT Project No 169125. \\

%\bibliography{bibliography}

\newpage

\begingroup
\squeezetable

\begin{table}
\caption{Phase coexistence and interfacial properties as a function of 
temperature for the ANC potential with different softness ($s$). All properties 
are dimensionless. }

\label{table-sup}

\begin{tabular}{ccccccc}
\hline\hline

\hspace{1.0cm}$s$\hspace{0.5cm} &\hspace{0.5cm} $T^{*}$\hspace{0.5cm} & 
\hspace{0.5cm} $\rho_{L}^{*}$\hspace{0.5cm} & \hspace{0.5cm} $\rho_{V}^{*}$ 
\hspace{0.5cm} & \hspace{0.5cm} $P^{*}$\hspace{0.5cm} & \hspace{0.5cm} 
$\gamma^{*}$\hspace{0.5cm} & \hspace{0.5cm} $B_2^{*}$\hspace{0.5cm} \\

\hline

0.2   & 0.3470  & 0.9456  & 0.1027  & 0.0246  & 0.0747  & -1.6186  \\ 
      & 0.3491  & 0.9288  & 0.1133  & 0.0270  & 0.0669  & -1.5825  \\
      & 0.3512  & 0.9141  & 0.1237  & 0.0292  & 0.0544  & -1.5471  \\
      & 0.3534  & 0.8948  & 0.1376  & 0.0309  & 0.0498  & -1.5124  \\
      & 0.3555  & 0.8715  & 0.1530  & 0.0329  & 0.0383  & -1.4783  \\
      & 0.3576  & 0.8429  & 0.1740  & 0.0353  & 0.0324  & -1.4449  \\
      & 0.3598  & 0.8044  & 0.2001  & 0.0378  & 0.0236  & -1.4121  \\
      & 0.3620  & 0.7683  & 0.2376  & 0.0408  & 0.0134  & -1.3799  \\

                                         \\ \hline

0.3   & 0.4000  & 0.9880  & 0.0573  & 0.0182  & 0.1499  & -1.9408 \\
      & 0.4048  & 0.9699  & 0.0653  & 0.0210  & 0.1338  & -1.8705 \\
      & 0.4097  & 0.9539  & 0.0742  & 0.0236  & 0.1129  & -1.8025 \\
      & 0.4146  & 0.9308  & 0.0877  & 0.0271  & 0.0995  & -1.7368 \\
      & 0.4196  & 0.9079  & 0.1012  & 0.0300  & 0.0743  & -1.6732 \\
      & 0.4247  & 0.8817  & 0.1173  & 0.0339  & 0.0613  & -1.6116 \\
      & 0.4298  & 0.8531  & 0.1407  & 0.0378  & 0.0442  & -1.5520 \\
      & 0.4350  & 0.8043  & 0.1687  & 0.0426  & 0.0317  & -1.4943 \\

                                           \\ \hline
0.4   & 0.4300  & 1.0417  & 0.0253  & 0.0099  & 0.2893  & -2.4651  \\
      & 0.4406  & 1.0195  & 0.0327  & 0.0128  & 0.2489  & -2.3089  \\
      & 0.4515  & 0.9956  & 0.0418  & 0.0162  & 0.2093  & -2.1624  \\
      & 0.4626  & 0.9669  & 0.0523  & 0.0202  & 0.1866  & -2.0248  \\
      & 0.4740  & 0.9344  & 0.0674  & 0.0253  & 0.1462  & -1.8954  \\
      & 0.4857  & 0.8971  & 0.0879  & 0.0315  & 0.1101  & -1.7736  \\
      & 0.4977  & 0.8535  & 0.1152  & 0.0390  & 0.0722  & -1.6588  \\
      & 0.5100  & 0.7934  & 0.1648  & 0.0483  & 0.0451  & -1.5504  \\
                                           \\ \hline

0.5   & 0.4900  & 1.0417  & 0.0253  & 0.0113  & 0.3157  & -2.4847 \\ 
      & 0.5026  & 1.0195  & 0.0327  & 0.0152  & 0.2800  & -2.3366 \\ 
      & 0.5154  & 0.9956  & 0.0418  & 0.0192  & 0.2395  & -2.1971 \\ 
      & 0.5287  & 0.9669  & 0.0523  & 0.0235  & 0.2075  & -2.0655 \\ 
      & 0.5422  & 0.9344  & 0.0674  & 0.0289  & 0.1686  & -1.9413 \\ 
      & 0.5561  & 0.8971  & 0.0879  & 0.0357  & 0.1306  & -1.8239 \\ 
      & 0.5704  & 0.8535  & 0.1152  & 0.0437  & 0.0955  & -1.7129 \\ 
      & 0.5850  & 0.7934  & 0.1508  & 0.0537  & 0.0572  & -1.6078 \\ 
                                             \\ \hline
                                           
\end{tabular}
\end{table}

\endgroup
                                             
\begingroup
\squeezetable

\begin{table}
\begin{tabular}{ccccccc}
\hline\hline

\hspace{1.0cm}$s$\hspace{0.5cm} &\hspace{0.5cm} $T^{*}$\hspace{0.5cm} & 
\hspace{0.5cm} $\rho_{L}^{*}$\hspace{0.5cm} & \hspace{0.5cm} $\rho_{V}^{*}$ 
\hspace{0.5cm} & \hspace{0.5cm} $P^{*}$\hspace{0.5cm} & \hspace{0.5cm} 
$\gamma^{*}$\hspace{0.5cm} & \hspace{0.5cm} $B_2^{*}$\hspace{0.5cm} \\

\hline

0.6   & 0.5500  & 1.0040  & 0.0267  & 0.0129  & 0.3636  & -2.5252 \\ 
      & 0.5645  & 0.9830  & 0.0334  & 0.0170  & 0.3148  & -2.3826 \\ 
      & 0.5794  & 0.9583  & 0.0423  & 0.0207  & 0.2852  & -2.2477 \\ 
      & 0.5947  & 0.9321  & 0.0537  & 0.0261  & 0.2380  & -2.1201 \\ 
      & 0.6104  & 0.9038  & 0.0655  & 0.0317  & 0.1926  & -1.9993 \\ 
      & 0.6265  & 0.8713  & 0.0842  & 0.0392  & 0.1524  & -1.8848 \\ 
      & 0.6430  & 0.8329  & 0.1061  & 0.0477  & 0.1163  & -1.7762 \\ 
      & 0.6600  & 0.7839  & 0.1412  & 0.0575  & 0.0803  & -1.6730 \\ 
                                              \\  \hline

0.7   & 0.6200  & 0.9840  & 0.0316  & 0.0166  & 0.3758  & -2.4991 \\ 
      & 0.6371  & 0.9627  & 0.0373  & 0.0212  & 0.3260  & -2.3617 \\ 
      & 0.6547  & 0.9382  & 0.0464  & 0.0261  & 0.2834  & -2.2316 \\ 
      & 0.6727  & 0.9123  & 0.0595  & 0.0323  & 0.2464  & -2.1082 \\ 
      & 0.6912  & 0.8831  & 0.0740  & 0.0391  & 0.1978  & -1.9911 \\ 
      & 0.7103  & 0.8438  & 0.0923  & 0.0477  & 0.1521  & -1.8798 \\ 
      & 0.7299  & 0.8043  & 0.1170  & 0.0582  & 0.1117  & -1.7741 \\ 
      & 0.7500  & 0.7472  & 0.1508  & 0.0706  & 0.0725  & -1.6734 \\ 
                         \\  \hline
                         
0.8   & 0.7000 & 0.9702  & 0.0357  & 0.0220  & 0.4019  & -2.4442 \\
      & 0.7197 & 0.9466  & 0.0437  & 0.0275  & 0.3515  & -2.3153 \\
      & 0.7399 & 0.9215  & 0.0536  & 0.0335  & 0.3020  & -2.1928 \\
      & 0.7607 & 0.8923  & 0.0672  & 0.0407  & 0.2457  & -2.0763 \\
      & 0.7821 & 0.8596  & 0.0850  & 0.0495  & 0.2041  & -1.9653 \\
      & 0.8041 & 0.8171  & 0.1033  & 0.0594  & 0.1506  & -1.8597 \\
      & 0.8267 & 0.7717  & 0.1331  & 0.0717  & 0.1005  & -1.7589 \\
      & 0.8500 & 0.7075  & 0.1740  & 0.0855  & 0.0594  & -1.6628 \\
                           \\  \hline  

0.9   & 0.7900  & 0.9573 & 0.0425  & 0.0287  & 0.4153  & -2.3831  \\
      & 0.8123  & 0.9331 & 0.0522  & 0.0355  & 0.3574  & -2.2636  \\
      & 0.8352  & 0.9069 & 0.0634  & 0.0430  & 0.3017  & -2.1496  \\
      & 0.8588  & 0.8762 & 0.0791  & 0.0519  & 0.2532  & -2.0408  \\
      & 0.8831  & 0.8411 & 0.0971  & 0.0622  & 0.1913  & -1.9369  \\
      & 0.9080  & 0.8010 & 0.1185  & 0.0747  & 0.1502  & -1.8376  \\
      & 0.9336  & 0.7491 & 0.1500  & 0.0883  & 0.0946  & -1.7427  \\
      & 0.9600  & 0.6857 & 0.1904  & 0.1051  & 0.0504  & -1.6518  \\
                                         \\  \hline

\end{tabular}
\end{table}

\endgroup

\begingroup
\squeezetable

\begin{table}[t]
\begin{tabular}{ccccccc}
\hline\hline

\hspace{1.0cm}$s$\hspace{0.5cm} &\hspace{0.5cm} $T^{*}$\hspace{0.5cm} & 
\hspace{0.5cm} $\rho_{L}^{*}$\hspace{0.5cm} & \hspace{0.5cm} $\rho_{V}^{*}$ 
\hspace{0.5cm} & \hspace{0.5cm} $P^{*}$\hspace{0.5cm} & \hspace{0.5cm} 
$\gamma^{*}$\hspace{0.5cm} & \hspace{0.5cm} $B_2^{*}$\hspace{0.5cm} \\

\hline

1.0   & 0.8500  & 0.9859  & 0.0370  & 0.0271  & 0.5245  & -2.5230  \\
      & 0.8761  & 0.9622  & 0.0454  & 0.0340  & 0.4635  & -2.3935  \\
      & 0.9029  & 0.9399  & 0.0561  & 0.0418  & 0.4007  & -2.2701  \\
      & 0.9306  & 0.9092  & 0.0691  & 0.0506  & 0.3313  & -2.1526  \\
      & 0.9591  & 0.8751  & 0.0815  & 0.0616  & 0.2691  & -2.0405  \\
      & 0.9885  & 0.8399  & 0.1043  & 0.0745  & 0.2097  & -1.9336  \\
      & 1.0188  & 0.7902  & 0.1340  & 0.0896  & 0.1502  & -1.8315  \\
      & 1.0500  & 0.7283  & 0.1677  & 0.1065  & 0.0899  & -1.7340  \\
                                            \\  \hline
\end{tabular}
\end{table}

\endgroup

\begingroup
\squeezetable

\end{document}